\documentclass[twocolumn]{aastex701}

\usepackage{amsmath}
\usepackage{amssymb}
\usepackage{xcolor}
\usepackage{placeins}

\newcommand{\uas}{\ensuremath{\mu\mathrm{as}}}
\newcommand{\Glam}{\ensuremath{\mathrm{G}\lambda}}
\newcommand{\fone}{\ensuremath{f_1}}
\newcommand{\fw}{\ensuremath{f_w}}
\newcommand{\mring}{\ensuremath{m}-ring}
\newcommand{\muas}{\ensuremath{\,\mu\mathrm{as}}}

\shorttitle{Submillimeter balloon VLBI prospects}
\shortauthors{Bagchi et al.}

\submitjournal{ApJ}

\begin{document}

\title{Prospects for submillimeter VLBI of black hole photon rings with a stratospheric balloon-borne telescope}

\author[gname=Mayukh,sname=Bagchi]{Mayukh Bagchi}
\affiliation{Department of Physics, Engineering Physics and Astronomy, Queen's University, Kingston, ON K7L 3N6, Canada}
\email{mayukh.bagchi@queensu.ca}

\author[gname=Dominic,sname=Pesce]{Dominic W. Pesce}
\affiliation{Harvard University, Cambridge, MA, USA}
\affiliation{Center for Astrophysics \textbar{} Harvard \& Smithsonian, Cambridge, MA, USA}
\email{dominic.pesce@cfa.harvard.edu}

\author[gname=Felix,sname=Thiel]{Felix M. Thiel}
\affiliation{Department of Physics, Engineering Physics and Astronomy, Queen's University, Kingston, ON K7L 3N6, Canada}
\email{21fmt2@queensu.ca}

\author[gname=Laura,sname=Fissel]{Laura M. Fissel}
\affiliation{Department of Physics, Engineering Physics and Astronomy, Queen's University, Kingston, ON K7L 3N6, Canada}
\email{laura.fissel@queensu.ca}

\author[gname=Lindy,sname=Blackburn]{Lindy Blackburn}
\affiliation{Harvard University, Cambridge, MA, USA}
\affiliation{Center for Astrophysics \textbar{} Harvard \& Smithsonian, Cambridge, MA, USA}
\email{lblackburn@cfa.harvard.edu}

\begin{abstract}
Submillimeter very long baseline interferometry (VLBI) achieves higher
angular resolution and probes more optically thin emission near supermassive black holes, with less interstellar scattering toward Sgr~A$^*$. Above
345\,GHz, however, the atmosphere confines ground-based VLBI to a few high,
dry sites. Using 2012--2022 weather, we forecast what a 2.5\,m
telescope floating at 35--40\,km, a successor to the 22\,GHz
Balloon-borne VLBI Experiment (BVEX) pathfinder, would add to ground-VLBI
arrays at 345--690\,GHz. The balloon adds little baseline length but
reaches higher frequencies, where a fixed baseline is longer in
wavelengths. On an ALMA--balloon baseline, Sgr~A$^*$ at 460\,GHz is at the
detection threshold, with ten or more 20\,s detections on about half
of the simulated April days. These detections lie beyond the polarimetric photon
ring transition, where the $n{=}1$ ring dominates the polarized
visibility amplitudes. M87$^*$ at 460\,GHz would need frequency phase transfer from a
230\,GHz reference to be detected. With phase transfer, most Sgr~A$^*$ detections
beyond the transition at 460\,GHz are on balloon baselines, whose
thermal noise meets the linear-polarization threshold within one
300\,s scan. At 690\,GHz such detections require phase transfer that
stays coherent at three times the reference frequency. In Fisher
forecasts, a 460\,GHz ring-geometry constraint requires balloons but
is weaker than at 345\,GHz, where balloons add little improvement. High-resolution 460\,GHz VLBI science would
require a 32\,GHz bandwidth backend, submillimeter receivers, and
simultaneous 230\,GHz reference recording at every station.
\end{abstract}


%% ============================================================================
%% SECTION 1: INTRODUCTION
%% ============================================================================
\section{Introduction} \label{sec:intro}

\begin{figure}[!htb]
\centering
\includegraphics[width=\columnwidth]{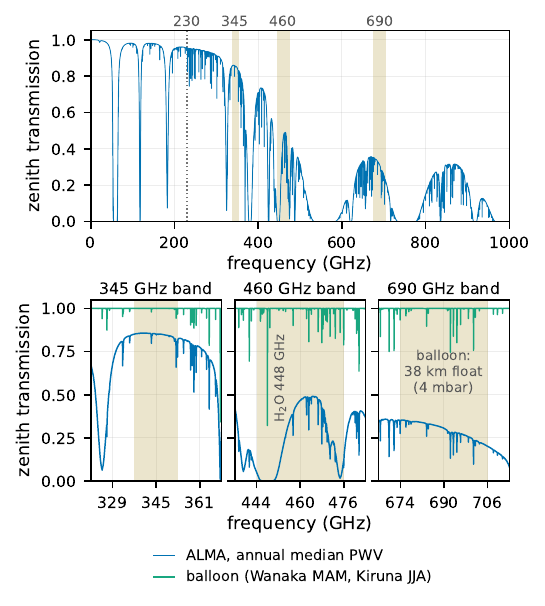}
\caption{Zenith atmospheric transmission from the \texttt{am} code
\citep{Paine2023_am} using MERRA-2 weather profiles
\citep{Gelaro2017_MERRA2}. Top panel: ALMA from 0 to 1000\,GHz for its
annual median water vapor content (1.0\,mm); the dotted line marks
230\,GHz, the reference frequency for frequency phase transfer
(Section~\ref{sec:synth}). Bottom panel: zoom-ins of the 345, 460, and
690\,GHz bands, comparing ALMA with a stratospheric balloon at a
38\,km float altitude above Wanaka, averaged over March to May (MAM).
The Kiruna transmission, averaged over June to August (JJA), differs
only in narrow line cores (Section~\ref{sec:bvex}). The shaded regions
indicate the bandpasses assumed in our simulations (16\,GHz at
345\,GHz, 32\,GHz at 460 and 690\,GHz). Within them the balloon
transmission is essentially unity, while the median ALMA transmission
is 0.85, 0.24, and 0.31, respectively. The 460\,GHz band contains the
448\,GHz water line, and the ALMA transmission at the band center is 0.47.}
\label{fig:atmtx}
\end{figure}

Balloon-borne telescopes operate at float altitudes of 35--40\,km,
above more than 99.5\% of the Earth's atmosphere, and can therefore
observe at submillimeter and far-infrared wavelengths. Such telescopes
could serve as mobile outrigger stations for ground-VLBI arrays, giving
higher angular resolution by permitting observations at frequencies
that are not accessible from most ground sites \citep{Pesce2024}
(Figure~\ref{fig:atmtx}). Two pathfinder experiments are attempting
to demonstrate that balloon telescopes can be used as VLBI stations.
BVEX (the Balloon-borne VLBI Experiment) is a 0.9\,m telescope at 22\,GHz, with the first
balloon-to-ground fringe detection as its primary goal. It flies on a Centre
National d'\'Etudes Spatiales (CNES) gondola launched through the Canadian
Space Agency (CSA) STRATOS program.
The first BVEX flight launched from Timmins, Ontario, in 2025 August,
but was unable to take VLBI data because damage to the launch balloon
meant that the experiment could not ascend to an altitude where the
telescope could be pointed \citep{Bagchi2026_SPIE,Thiel2026_SPIE}.
Another balloon VLBI effort is the 20\,GHz system of
\citet{Doi2019_balloonVLBI}, which JAXA flew as a technology test to a
26\,km float on 2026 June 14 \citep{ISAS2026_B26_02}.

A black hole surrounded by optically thin emission produces a stack of
lensed sub-images, the direct $n{=}0$ image and a series of thinner
photon rings with $n \geq 1$ \citep{Johnson2020}. Two measurements of
the $n{=}1$ ring are within reach of ground arrays that include a
balloon. The first is its diameter, which tracks the mass-to-distance
ratio with only a weak dependence on the emitting plasma
\citep{Johnson2020}. The Event Horizon Telescope (EHT) images of
M87$^*$ and Sgr~A$^*$ \citep{EHTC2019_I,EHTC2022_I} are dominated by
the direct emission, which blends with the ring at 230\,GHz resolution
\citep{Johnson2020}. The second measurement needs no image, because
beyond a transition baseline, which we call the polarimetric
transition, the $n{=}1$ ring dominates the polarized
visibility and shifts the phase of an interferometric polarization
quotient \citep{Palumbo2023_polarimetry}.

Unlike a balloon, space VLBI, such as the proposed Black Hole Explorer
(BHEX; \citealt{Johnson2024_BHEX}), lengthens the baselines
beyond the Earth's diameter. At higher frequency the emission near the
black hole is also more optically thin \citep{Falcke2000}, and the
interstellar scattering toward Sgr~A$^*$ decreases as the square of the
wavelength \citep{Johnson2018_scattering}. A fixed baseline between the
Atacama Large Millimeter/submillimeter Array (ALMA) and a balloon is
longer in wavelengths at higher frequency.
A longer flight mostly repeats the same baselines rather than lengthening them,
since a balloon launched from Wanaka, New Zealand, circles the Earth about once every
7--10 days.

On a given physical baseline, a higher observing frequency increases
the photon ring contribution to the visibilities. A higher frequency
also lowers the signal-to-noise ratio, because the long-baseline
signals are much fainter and the system noise is higher, both from the
receivers and from the emission of the more opaque atmosphere above
the ground stations. At 345\,GHz and below, fringes are
easy to detect but the photon ring contributes little structure on the
available Earth-size baselines. At 690\,GHz the ring dominates the long-baseline
visibilities \citep{Johnson2020}, but fringe detection is limited by
sensitivity.

Longer integrations improve sensitivity only while the phase stays
coherent. EHT stations at 230\,GHz integrate coherently for about
10\,s \citep{EHTC2019_II,EHTC2019_III}. Frequency phase transfer (FPT) can extend this coherence
time by measuring the phase in a simultaneously
observed lower-frequency reference band and applying it, scaled by the
frequency ratio, to the science band
\citep{RiojaDodson2011,Rioja2023_FPT}. Using an 86\,GHz reference, FPT
has extended the coherence at 215 and 258\,GHz on Earth-size baselines
\citep{Issaoun2025_FPT,Zhao2025_FPT}.

This paper investigates what a submillimeter successor to the
balloon VLBI pathfinders described above could add to the ground-VLBI arrays
of the next decade. We forecast how far balloon stations could improve
the ability of these arrays to constrain models of the photon ring
emission. We generate synthetic observations with historical weather
using versions of \texttt{ngehtsim} \citep{Pesce2024} and
\texttt{eht-imaging} modified to support balloon stations. Section~\ref{sec:platforms}
describes the balloon platform, the ground arrays, and the simulation
pipeline. Section~\ref{sec:detect} tests fringe detection on the
single ALMA--balloon baseline and the coverage of the full arrays with
FPT.
Section~\ref{sec:geometry} provides estimated reconstruction
uncertainties for a geometric parametrization of the photon ring at
345 and 460\,GHz, based on a Fisher analysis with
\texttt{ngEHTforecast}\footnote{\url{https://github.com/aeb/ngEHTforecast}}.
Section~\ref{sec:discussion} discusses the polarimetric regime, the
limitations, and the outlook, and Section~\ref{sec:conclusions}
presents our conclusions. A companion paper \citep{Thiel2026_jet} uses the same modified
software at 86 and 230\,GHz, where balloon baselines from Kiruna, Sweden, fill
short $(u,v)$ spacings and improve images of the extended jet of
M87$^*$. That paper addresses dense $(u,v)$ coverage at millimeter
wavelengths, while this one focuses on the detectability of the photon
ring at submillimeter wavelengths.

%% ============================================================================
%% SECTION 2: PLATFORMS AND SYNTHETIC OBSERVATIONS
%% ============================================================================
\section{Observing Platforms and Synthetic Observations} \label{sec:platforms}

\subsection{submm-BVEX: A Stratospheric VLBI Station} \label{sec:bvex}

This paper forecasts a submillimeter successor to the 0.9\,m,
22\,GHz BVEX pathfinder introduced in
Section~\ref{sec:intro}. We call the successor submm-BVEX to
distinguish it from the pathfinder, and for this study we assume it to
be a 2.5\,m dish that operates at 35--40\,km above sea level.

\begin{figure*}[!htb]
\centering
\includegraphics[width=0.35\textwidth]{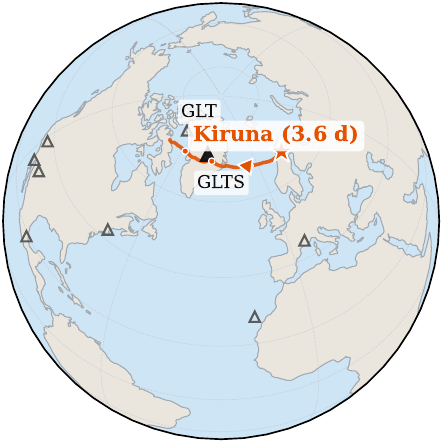}\hspace{0.04\textwidth}%
\includegraphics[width=0.35\textwidth]{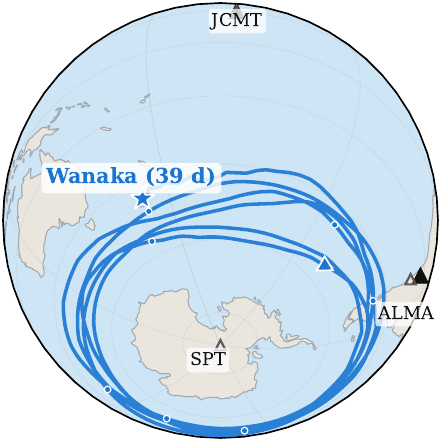}
\caption{Recorded balloon trajectories used in this paper: Kiruna (left,
3.6\,d, orange; the M87$^*$ single-baseline trajectory, from the 2024 TRANSAT flight, with
data provided by CNES and the CSA) and Wanaka (right,
39\,d, blue; the SuperBIT flight, \citealt{Gill2024_SuperBIT}). Stars
mark launches, dots flight days, and the small filled arrowhead in each
trajectory color marks the drift direction. Filled triangles are the
460\,GHz stations and open triangles the other ngEHT 345\,GHz
stations. The two-balloon campaigns of both sources assume the same Wanaka
trajectory with the launches staggered by 16\,d
(Section~\ref{sec:access}).}
\label{fig:trajectories}
\end{figure*}

\begin{table*}[!htb]
\centering
\footnotesize
\caption{Receiver assumptions for the simulated 460\,GHz (Band 8)
and 690\,GHz (Band 9) stations.}
\label{tab:stations}
\begin{tabular}{lccccl}
\hline\hline
Station & \multicolumn{2}{c}{Band 8 (460\,GHz)} & \multicolumn{2}{c}{Band 9 (690\,GHz)} & References \\
 & $T_R$ (K) & sideband & $T_R$ (K) & sideband & \\
\hline
ALMA & 120$^{a}$ & 2SB (SSR $\approx 0.1$) & 80 & DSB (SSR $= 1$) & \citealt{Carpenter2023_WSU}; \citealt{Baryshev2015} \\
APEX & 150 & 2SB (SSR $\approx 0.1$) & 225$^{b}$ & 2SB (SSR $\approx 0.1$) & \citealt{APEX_nFLASH}; \citealt{APEX_SEPIA660} \\
JCMT & 200$^{c}$ & 2SB & 150 & 2SB & \citealt{Bintley2024} \\
GLTS & 200$^{c}$ & 2SB & 150$^{d}$ & 2SB & \citealt{Bintley2024} \\
submm-BVEX$^{e}$ & 120 & 2SB (SSR $\approx 0.1$) & 110 & DSB (SSR $= 1$) & this work (Section~\ref{sec:bvex}) \\
\hline
\end{tabular}

\begin{minipage}{\textwidth}
\vspace{4pt}\footnotesize
$^{a}$WSU Band 8 goal (Fig.~34 of \citealt{Carpenter2023_WSU}); the
qualification cartridge measured 100\,K at band center
\citep{Satou2008}.\\
$^{b}$SEPIA660, commissioned in 2018
\citep{MontenegroMontes2019}, with about 225\,K at 600\,GHz and below
335\,K across the band.\\
$^{c}$Aspirational; no
460\,GHz VLBI receiver exists at either site.\\
$^{d}$Assumes a Kuntur-class cartridge at the GLTS
\citep{Bintley2024}.\\
$^{e}$Forecast configuration of
Section~\ref{sec:bvex}: 50\,K at 345\,GHz is at the low end of the
45--75\,K measured for the ALMA Band 7 (275--373\,GHz) cartridges
\citep{Mahieu2012}, and we assume the same at 230\,GHz; 120\,K is the
Band 8 goal of note $a$; 110\,K DSB lies in the 75--125\,K range of
the ALMA Band 9 cartridges \citep{Baryshev2015,Belitsky2018}. ALMA's
80\,K at 690\,GHz is our assumption at the low end of that range;
\citet{Pesce2024} adopt 100\,K, which would raise the ALMA--balloon
noise at 690\,GHz by a few percent.\\
Notes: 2SB and DSB are two-sideband and
double-sideband receivers; SSR is the sideband separation ratio, and
$T_\mathrm{sys}$ scales as $(1 + \mathrm{SSR})$ \citep{Pesce2024}. We assume that ALMA is operated in phased-array mode and that all
stations record 32\,GHz in both bands. The JCMT is 15\,m and the GLTS
12\,m (omitted for Sgr~A$^*$).
\end{minipage}
\end{table*}

The residual atmosphere at the float altitudes for the balloon is thin
enough that we can assume no atmospheric contribution at all, so
opacity, sky brightness, and wind loading do not enter the balloon SEFD
(system equivalent flux density). At a 38\,km float altitude the MERRA-2
zonal-average profiles for the latitude bands of Wanaka (averaged over
March to May) and Kiruna (averaged over June to August) predict a
0.16--0.17\,$\mu$m water vapor column, compared to a typical 1\,mm
column at ALMA. The transmission spectrum is essentially identical from
different stratospheric locations, varying only in the cores of narrow
atmospheric lines. The
SEFD for the balloon telescope is set by the dish diameter, the receiver
temperature and sideband configuration, and the aperture efficiency.

We assume the submm-BVEX receiver temperatures listed in
Table~\ref{tab:stations}, each tied to demonstrated or funded hardware
in its note~$e$. Every parameter we assume for
submm-BVEX is a design target, including these receiver temperatures, since
no submillimeter hardware has yet been built for BVEX. We note, however, that
superconducting heterodyne receivers have previously
flown on balloons. For example, TELIS carried a 480--650\,GHz
superconductor--insulator--superconductor (SIS) receiver, with a
pre-flight DSB noise temperature as low as 120\,K, in a liquid-helium
cryostat on a 2009 Kiruna flight \citep{deLange2010}. We assume
a nominal aperture efficiency of 0.70, which our assumed 10\,$\mu$m rms
surface accuracy reduces to 0.69, 0.67, and 0.64 at 345, 460, and
690\,GHz, respectively. We also assume no efficiency loss due to pointing errors, since the sub-arcsecond gondola
pointing of SuperBIT \citep{Gill2024_SuperBIT} is well within the
roughly 40\arcsec{} beam FWHM of a 2.5\,m dish at 690\,GHz. With these values, the
projected balloon antenna SEFD is about 45\,kJy at 345\,GHz, 113\,kJy at 460\,GHz, and
197\,kJy at 690\,GHz.

In terms of observation bandwidth, we assume 16\,GHz at 230
and 345\,GHz, the bandwidth per polarization of the reference array of
the next-generation EHT (ngEHT; \citealt{Doeleman2023_ngEHT}), 8\,GHz in
each of two sidebands, and 32\,GHz at 460 and 690\,GHz, with submm-BVEX
recording the same bandwidth as its ground-station partners. In every
campaign we assume that all stations simultaneously record 230\,GHz at
the science bandwidth to enable FPT, as
described in Section~\ref{sec:synth}. The 32\,GHz bandwidth is
the goal of the ALMA Wideband Sensitivity Upgrade (WSU;
\citealt{Carpenter2023_WSU}). Assuming two-bit sampling and two polarizations,
32\,GHz of recorded bandwidth is a data rate of 256\,Gbit\,s$^{-1}$,
about thirty times the 8.2\,Gbit\,s$^{-1}$ of the pathfinder BVEX backend
\citep{Bagchi2026_SPIE}.

The frequency standard for submm-BVEX is assumed to be an oven-controlled crystal oscillator
(OCXO), since mass and power rule out a hydrogen maser on a balloon.
For ground--ground baselines we assume coherent integrations of 10\,s,
which is standard EHT practice at 230\,GHz
\citep{EHTC2019_II,EHTC2019_III}. On ground--balloon baselines we
assume 5\,s, with the balloon coherence time set by the clock stability and
position-tracking precision rather than by the atmosphere. We apply both
values at every observing frequency. The $\lambda/2\pi$
position-tracking requirement that \citet{Thiel2026_SPIE} set for BVEX
at 22\,GHz becomes about 100\,$\mu$m at 460\,GHz. A 5\,s coherence at 460\,GHz
requires a fractional frequency stability near $7 \times 10^{-14}$ over
the integration. An oscillator that meets this requirement at 460\,GHz
supports only about 3.3\,s at
690\,GHz, so the 5\,s we assume there is optimistic by a factor of 1.5,
which is the ratio of the two frequencies. Single-baseline detections must therefore be achieved within
a few seconds, and longer integrations require the
FPT of Section~\ref{sec:synth}.

The balloon trajectory is set by the launch site and the stratospheric
winds, which at float altitude blow mainly along lines of constant
latitude and carry a balloon east-west around its launch latitude. Balloon telescopes cannot currently be steered. Launch sites are few, so a balloon team must pick a location that
will optimize the $(u,v)$ coverage on event-horizon scales. We choose
for each source a single-baseline launch site from which the balloon
and ALMA can see the source together for several hours per day. For
M87$^*$ ($+12.4\arcdeg$ declination) we use a flight from Kiruna,
Sweden (lat $= +67\arcdeg$), for which both the balloon telescope and
ALMA can see the source about 6 hours per day. For Sgr~A$^*$
($-29\arcdeg$ declination) we use a flight from Wanaka, New Zealand
(lat $= -45\arcdeg$), along which ALMA and submm-BVEX can both see
Sgr~A$^*$ for 4.8 hours per day in April. The ten-day, two-balloon
campaigns of both sources use two launches from Wanaka on its recorded
39\,d trajectory (Section~\ref{sec:access}).
Figure~\ref{fig:trajectories} shows the two recorded trajectories as well as the locations of
the ground VLBI stations with which they can potentially correlate.

\begin{figure*}[!htb]
\centering
\includegraphics[width=\textwidth]{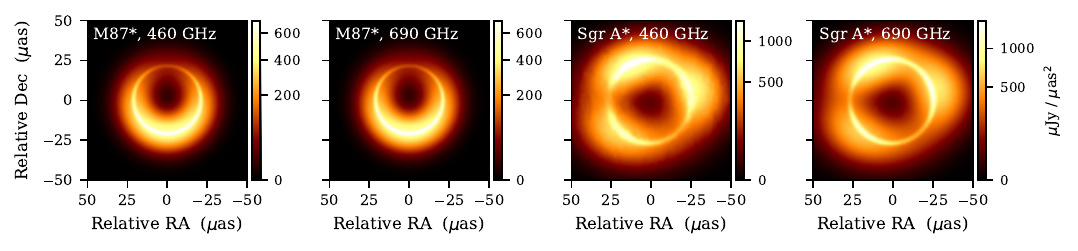}
\caption{Input intensity models used in this work, taken from Appendix F of \citet{Pesce2024}.
\textit{Left panels:} Input models for M87$^*$ simulations at 460 and
690\,GHz. Note that we assume that the intensity distribution does not
vary with frequency for these bands. \textit{Right panels:} Input
intensity maps for Sgr~A$^*$ at 460 and 690\,GHz. The input intensity
distributions are the same for both bands, but both models include
frequency-dependent scattering. All panels use a square-root color
scale over a 100\muas{} field. Total input model fluxes are
0.600\,Jy for M87$^*$ and 2.5\,Jy for Sgr~A$^*$. These input models are used to generate the forecasting simulations described in
Section~\ref{sec:detect}.}
\label{fig:models}
\end{figure*}

\subsection{Ground Arrays at 345, 460, and 690\texorpdfstring{\,}{ }GHz} \label{sec:ground}

At 345\,GHz the benchmark is the 14-station 345\,GHz configuration of
\texttt{ngehtsim} \citep{Pesce2024}, based on the ngEHT reference
array \citep{Doeleman2023_ngEHT}, with science goals set out in
\citet{Johnson2023_ngEHT}. We treat this array as the future ground array, and investigate how balloon telescopes could add to it. The South Pole Telescope (SPT) never sees
M87$^*$ and the Greenland Telescope never sees Sgr~A$^*$, so each source
is observed by 13 of the 14 stations. At the three new sites (San Pedro
M\'{a}rtir Observatory in Mexico, Teide Observatory in the Canary
Islands, and Las Campanas Observatory in Chile) we assume
10\,m dishes, the \texttt{ngehtsim} default for a new site.

At 460\,GHz no comparable array forecasts exist, so we choose the sites whose atmospheric conditions allow VLBI at these frequencies. For ground-based stations in our 460 and 690\,GHz simulations, we include ALMA, the Atacama Pathfinder
Experiment (APEX), the James Clerk Maxwell Telescope (JCMT), and the
Greenland Telescope at Summit Camp (GLTS). These sites are few, with
ALMA and APEX co-located, so long 460\,GHz ground baselines are scarce. System temperatures
and SEFDs are calculated following the methodology of \citet{Pesce2024}, from the
receiver temperatures, bandwidths, and sideband types listed in Table~\ref{tab:stations}.
Our simulated 690\,GHz array consists of the same four stations with Band 9 (690\,GHz) receivers.

\subsection{Generation of Synthetic Observations} \label{sec:synth}

\begin{figure*}[!htb]
\centering
\includegraphics[width=\textwidth]{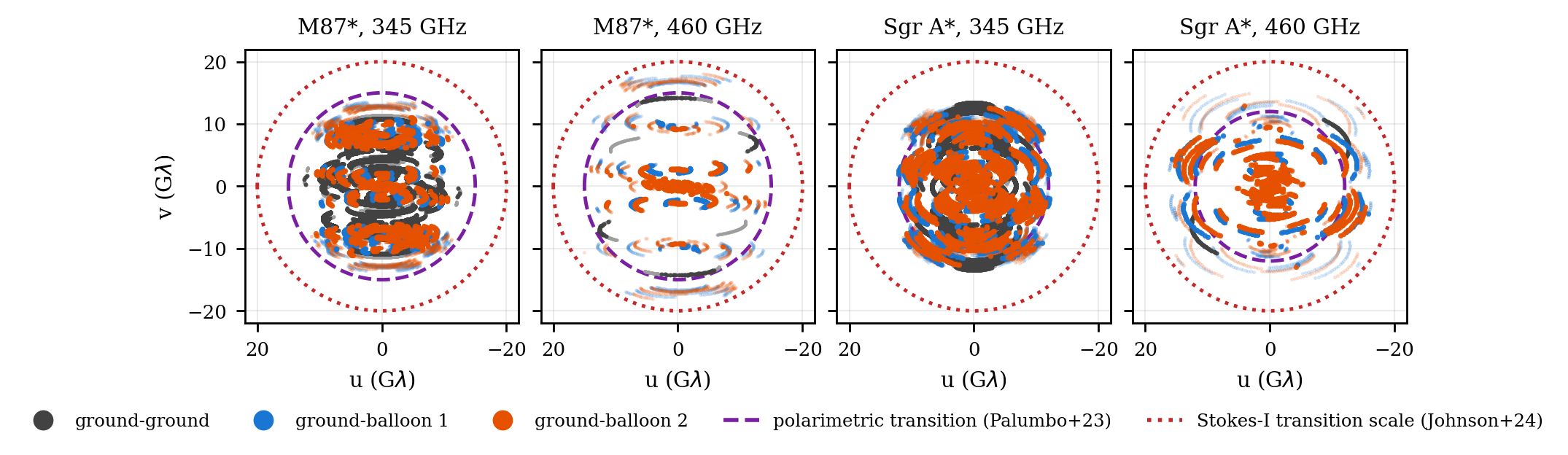}
\caption{$(u,v)$ coverage of the ten-day two-balloon campaigns, from
left: M87$^*$ at 345 and 460\,GHz, then Sgr~A$^*$ at 345 and 460\,GHz,
each with two staggered Wanaka launches, on the ngEHT array at
345\,GHz and on ALMA, APEX, JCMT, and GLTS at 460\,GHz. Gray points
are ground--ground baselines, blue and orange points the baselines to
the first and second balloon. Faint points are every scheduled scan;
solid points pass the detection rule of Section~\ref{sec:single} on
300\,s phase-transferred scans. Dashed circle: polarimetric
transition (15 and 12\,\Glam; \citealt{Palumbo2023_polarimetry});
dotted circle: Stokes-I transition scale (about 20\,\Glam;
\citealt{Johnson2024_BHEX}).}
\label{fig:uvcoverage}
\end{figure*}

Each forecast starts by generating a synthetic observation using
\texttt{ngehtsim}, following the methodology of \citet{Pesce2024}. For
each ground site and observing day the code reconstructs the atmospheric
opacity and brightness temperature from historical weather data. It
calculates the station SEFD from the dish diameter, aperture efficiency,
receiver temperature, and the calculated atmospheric opacity, and it
draws per-baseline thermal noise from the radiometer equation. The code
evaluates the opacity and sky brightness at the center of the recorded
band. At 460\,GHz the 32\,GHz band (444--476\,GHz) contains the
448\,GHz water line (Figure~\ref{fig:atmtx}). We therefore calculated
a band-averaged ground-station SEFD outside the simulation, weighting
each frequency channel by its inverse noise variance, so the
transparent part of the band dominates and the opaque line core adds
little. The band average raises the thermal noise by a factor
of about 1.5 on ALMA--balloon baselines and 1.6 on ground--ground
baselines. At 345 and 690\,GHz the ALMA--balloon noise changes by less
than 3\%. We keep the band-center values in the simulations and give the
band-averaged results alongside them at 460\,GHz. 

Neither \texttt{ngehtsim} nor the underlying \texttt{eht-imaging} library
\citep{Chael2016,Chael2018} supports balloon stations, so we modified
both software packages. In our version the recorded flight trajectory (time, latitude,
longitude, and altitude) is interpolated to each timestamp and
converted to Earth-centered coordinates before the $(u,v)$
calculation, so the synthetic baselines follow the actual motion of
the balloon telescope. The modified code also bypasses the atmospheric opacity,
sky brightness, and wind-loading terms for balloon stations
and applies separate coherence times to ground--ground and
ground--balloon baselines in the fringe-detection and
FPT analysis described below. These modifications, developed
jointly for this project and the millimeter balloon-VLBI analysis presented in \citet{Thiel2026_jet}, allow balloon station forecasts to be generated by the same software as the forecasts for ground stations.

The input source intensity models (Figure~\ref{fig:models}) are the geometric models
\citep[][see their Appendix F]{Pesce2024}, sampled at the observed
$(u,v)$ points. Each model is a pair of concentric
rings with a diameter of 42\,\uas{} for M87$^*$ and
52\,\uas{} for Sgr~A$^*$. A wide ring represents the direct emission
and carries 90\% of the flux, and a thin ring represents the photon
ring and carries 10\%. The M87$^*$ model is the same at every
frequency we simulate (345, 460, and 690\,GHz, and the 230\,GHz
FPT reference). The Sgr~A$^*$ model has an elongation of 1.2 and is
scattered at each frequency with the screen described in
\citet{Johnson2018_scattering}.

The simulated observing campaigns take place in April, when the EHT
typically observes and balloon launches take place from Wanaka, NZ.
The ten-day campaigns use the reconstructed weather of 2020 April
15--24, while the Monte Carlo single-baseline detectability analysis
of Section~\ref{sec:single} draws atmospheric opacities for each day
from the April weather of 2012 to 2022. The weather in
\texttt{ngehtsim} is a daily average of the MERRA-2 reanalysis, which
is sampled every 3 hours \citep{Pesce2024}, so it is constant within
each simulated day.

Our simulations reach integration times longer than the coherence
times of Section~\ref{sec:bvex} by using FPT. FPT is the first of the
two calibration steps of source-frequency phase referencing
\citep{RiojaDodson2011,Rioja2023_FPT}. \citet{Rioja2023_FPT} and
\citet{Issaoun2023_86GHz} proposed it for the ngEHT bands up to
345\,GHz, and we also apply it at 460 and 690\,GHz. The phase of a
230\,GHz reference fringe observed at the same time is solved every
5\,s on ground--balloon and every 10\,s on ground--ground baselines, and the successive solutions are joined into a continuous phase
solution (phase connection). That phase is multiplied by the
frequency ratio and subtracted from the higher-frequency band phase, and the
phase-corrected data are integrated over 300\,s. We call each
coherently averaged 5 or 10\,s data point an integration and a 300\,s
block of integrations a scan. Coherent
averaging sums the complex visibilities, so the noise falls as the
square root of the averaging time, but only while the phase is stable.
Incoherent averaging sums the visibility amplitudes or powers, which
needs no phase stability but lowers the noise more slowly. The
simulation keeps a scan when its 230\,GHz reference fringe reaches a
signal-to-noise ratio of 5 in one integration. The simulation then treats the 300\,s average of the
phase-corrected science-band data as coherent. In
Section~\ref{sec:access} we test this assumption. An FPT-corrected VLBI
system could allow far longer scans than 300\,s; we chose
300\,s as a compromise between integration depth and the known intra-scan
variability of Sgr~A$^*$ \citep{EHTC2022_IV}. FPT sets a
receiver requirement at both ends of the baseline, since the balloon
and every ground station must simultaneously record the reference band and the science
band. ALMA and APEX currently cannot record a 230\,GHz reference
simultaneously with a submillimeter band. FPT has
not yet been demonstrated for a balloon telescope, but could be
demonstrated by mm-BVEX, a planned millimeter successor to BVEX
\citep{Thiel2026_jet}.

Figure~\ref{fig:uvcoverage} shows the resulting $(u,v)$ coverage predicted
for the ten-day two-balloon campaigns at 345 and
460\,GHz. At
345\,GHz the balloons add tracks at and beyond the longest ngEHT ground
baselines, and at 460\,GHz, where the ground sites are few, they provide
most of the long-baseline coverage.

%% ============================================================================
%% SECTION 3: DETECTABILITY AND COVERAGE
%% ============================================================================
\section{Fringe Detectability and Baseline Coverage} \label{sec:detect}

\subsection{Single-baseline detectability on ALMA--submm-BVEX} \label{sec:single}

\begin{figure*}[!htb]
\centering
\includegraphics[width=\textwidth]{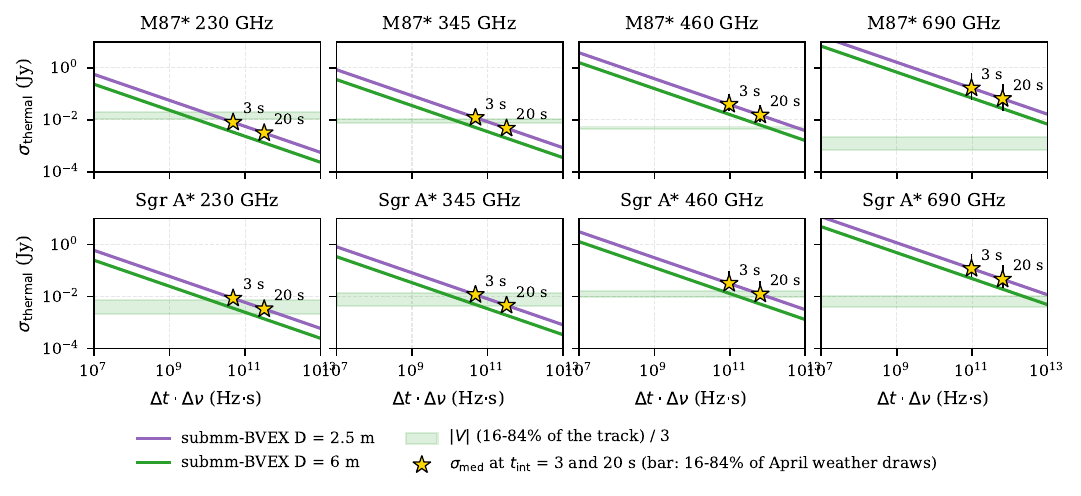}
\caption{Thermal noise level on the ALMA--submm-BVEX baseline against the
time--bandwidth product in April weather, for M87$^*$ (top) and
Sgr~A$^*$ (bottom) at 230, 345, 460, and 690\,GHz. The purple line is
the thermal noise for the fiducial 2.5\,m submm-BVEX dish and the green line
for a 6\,m dish; the green band is the visibility amplitude divided by 3 over the 16th--84th
percentile of the track amplitude, so a dish line below the band meets
the $3\sigma$ criterion for those integrations. Gold stars are the
simulated 2.5\,m noise at 3 and 20\,s at the recorded bandwidth of
each band, with bars spanning the 16--84\% range of the weather draws.
Table~\ref{tab:treq} lists the Sgr~A$^*$ 460\,GHz integration time requirements from
this analysis.}
\label{fig:sb_feasibility}
\end{figure*}

We first ask whether a fringe can be detected on ALMA--submm-BVEX,
the balloon baseline with the lowest thermal noise in the array, since
ALMA has by far the lowest SEFD of any station. We use a
minimal set of assumptions: no FPT, no fringe
grouping (in which fringe solutions on the strong baselines to a
station are used to calibrate its weaker baselines), and no coherent
or incoherent averaging across integrations. Every integration is
assessed based on its thermal noise $\sigma$, calculated from the
radiometer equation for the weather drawn in the Monte Carlo described
below. We count an integration as detected only if the observed
visibility amplitude is $>3\sigma$ and the noiseless model visibility
amplitude, sampled at the same $(u,v)$ point, also exceeds $3\sigma$.
The latter requirement is included because the amplitude of
a pure-noise sample is Rayleigh distributed, so a $3\sigma$ amplitude
cut alone passes noise 1.1\% of the time, which would result in
multiple false detections per day at our sampling rate. We call this
two-condition test the detection rule. A search for a fringe by
varying the delay and rate requires a signal-to-noise ratio near 5 to
7 \citep{Blackburn2019_HOPS}. Incoherent averaging of the integrations
in a scan can reach this level if each integration has a
signal-to-noise ratio of about 1 or more and the scan holds a few tens
of integrations \citep{Rogers1995}. Once the
fringe search has found the delay and rate, tracking the fringe phase
requires a detection above $3\sigma$ within one integration. We apply
the $3\sigma$ detection rule to each integration without incoherent
averaging over a scan, so the fractions of integrations detected that
are presented in this subsection are conservative.

We assume the telescope and receiver configuration for submm-BVEX
described in Section~\ref{sec:bvex}, a 2.5\,m dish with the per-band
receiver temperatures and recorded bandwidths listed in
Table~\ref{tab:stations}, combined with a phased ALMA
\citep{Goddi2019}, which we take to be equivalent to a 75\,m
single-dish aperture following \citet{Pesce2024}. We simulate
observations for each source for one 24-hour day, sampling every 63\,s
(a 3\,s integration followed by a 60\,s gap), and derive the noise at
10 and 20\,s from the 3\,s schedule by the radiometer scaling
$\sigma \propto 1/\sqrt{t_\mathrm{int}}$. This gives 336 scheduled
integrations per day for M87$^*$, where ALMA is paired with a balloon
on the Kiruna trajectory, and 274 for Sgr~A$^*$, where ALMA is paired
with a balloon on the Wanaka trajectory; in both cases the balloon
follows the first 24 hours of its recorded flight. Every fraction in
this subsection is a fraction of these scheduled integrations. We use a Monte Carlo that draws the ALMA
opacity and sky temperature calculated from a random day in April in a
random year between 2012 and 2022, with the source elevation profile
and $(u,v)$ track of 2020 April 15 held fixed. Each draw gives the
baseline noise versus time for one day. We use 100 draws at 460 and
690\,GHz and 20 draws at 230 and 345\,GHz, where the noise varies less
with weather (by a factor of 1.2--1.6 between the 16th and 84th
percentile draws, against 4--9 at 460 and 690\,GHz). The 10 and
20\,s integrations exceed the roughly 5\,s coherence time that we
expect submm-BVEX's OCXO and position tracking precision to support,
but we include them to show how the noise scales with integration
time.

\begin{table}[!htb]
\centering
\footnotesize
\caption{Single-baseline detection fractions for ALMA--submm-BVEX in April
weather.}
\label{tab:fractions}
\begin{tabular}{lcccc}
\hline\hline
 & \multicolumn{2}{c}{$t_\mathrm{int} = 10$\,s} & \multicolumn{2}{c}{$t_\mathrm{int} = 20$\,s} \\
Source, $\nu$ (GHz) & median & days $\geq 10$ & median & days $\geq 10$ \\
\hline
M87$^*$, 230 & 100\% & 100\% & 100\% & 100\% \\
M87$^*$, 345 & 73\% & 100\% & 98\% & 100\% \\
M87$^*$, 460 & 0\% & 0\% & 0\% & 0\% \\
M87$^*$, 690 & 0\% & 0\% & 0\% & 0\% \\
Sgr~A$^*$, 230 & 51\% & 100\% & 68\% & 100\% \\
Sgr~A$^*$, 345 & 56\% & 100\% & 76\% & 100\% \\
Sgr~A$^*$, 460 & 15\% & 58\% & 35\% & 72\% \\
Sgr~A$^*$, 460$^{a}$ & 0\% & 21\% & 9\% & 55\% \\
Sgr~A$^*$, 690 & 0\% & 0\% & 0\% & 0\% \\
\hline
\end{tabular}

\begin{minipage}{0.95\columnwidth}
\vspace{4pt}\footnotesize
Median daily fraction of the scheduled integrations (336 for M87$^*$
on Kiruna, 274 for Sgr~A$^*$ on Wanaka) that pass the detection rule,
and the percentage of April weather draws with at least ten
integrations that pass the 3$\sigma$ detection rule, from 100 draws at
460 and 690\,GHz and 20 draws at 230 and 345\,GHz. $^{a}$With the
460\,GHz ground-station noise averaged over the recorded band
(Section~\ref{sec:synth}).
\end{minipage}
\end{table}

Table~\ref{tab:fractions} lists, for each source, frequency, and
integration time, the median daily detection fraction and the
percentage of April days sampled that have ten or more integration
detections. In the
230\,GHz reference band M87$^*$ is detected on every 10\,s integration
and Sgr~A$^*$ on about half, so the fringe detection at the reference
band needed for FPT is present on the
ALMA--balloon baseline. At 345\,GHz
both sources are detected on more than half of their 10\,s
integrations. At 230 and 345\,GHz the Wanaka trajectory samples the
scattered Sgr~A$^*$ image at 6--11\,\Glam, where the visibility
amplitude is only a few times the 10\,s noise. Sgr~A$^*$ at 460\,GHz
is at threshold, and its detections fall on the shorter baselines of
the trajectory, 12.1--13.9\,\Glam, where the visibility amplitude is
higher. Since we only simulate the trajectory for the first day of the
balloon flight, the baseline variation is mostly due to Earth-rotation
aperture synthesis. With the ground-station
noise averaged over the recorded band (see Section~\ref{sec:synth}), ten or
more 20\,s detections occur at 460\,GHz on about half of the sampled
April days. The ALMA--balloon baseline at
460\,GHz runs from 12.0 to 14.2\,\Glam{} in this 24-hour simulation, so
these detections lie
beyond the 12\,\Glam{} polarimetric transition of Sgr~A$^*$
discussed in Section~\ref{sec:access}. M87$^*$ at 460\,GHz is not detected in any
of the 100 April weather draws, because its median visibility amplitude
of 14\,mJy lies below the 21\,mJy median 10\,s baseline noise. At
690\,GHz neither source is detected at any integration time. Launching
in June to August, the driest months at the Chilean sites, would raise
the median daily Sgr~A$^*$ fraction at 460\,GHz for 10\,s integrations
(band-center noise) from 15\% to 24\%, but balloon launches do not
take place from Wanaka during those months.

\begin{table}[!htb]
\centering
\footnotesize
\caption{Integration time required for a $3\sigma$ detection of the median
Sgr~A$^*$ visibility amplitude at 460\,GHz on ALMA--submm-BVEX in April median weather.}
\label{tab:treq}
\begin{tabular}{lcccc}
\hline\hline
Bandwidth & $D=2$\,m & $D=2.5$\,m & $D=3$\,m & $D=6$\,m \\
\hline
16\,GHz & 61.9\,s & 39.6\,s & 27.5\,s & 6.9\,s \\
32\,GHz (fiducial) & 31.0\,s & 19.8\,s & 13.8\,s & 3.4\,s \\
64\,GHz & 15.5\,s & 9.9\,s & 6.9\,s & 1.7\,s \\
\hline
\end{tabular}

\begin{minipage}{0.95\columnwidth}
\vspace{4pt}\footnotesize
Integration time at which the median thermal noise of the draws equals a
third of the median track amplitude. Each bandwidth doubling halves
every integration time; 16\,GHz is the WSU first step, 32\,GHz its goal, and
64\,GHz a further doubling beyond the WSU \citep{Carpenter2023_WSU}.
Averaging the ALMA noise over the entire recorded bandwidth, which at
32 and 64\,GHz includes the 448\,GHz water line, raises the integration time
estimates by factors of about 1.4, 2.3, and 1.7 at 16, 32, and
64\,GHz (Section~\ref{sec:synth}).
\end{minipage}
\end{table}

\begin{figure*}[!htb]
\centering
\includegraphics[width=\textwidth]{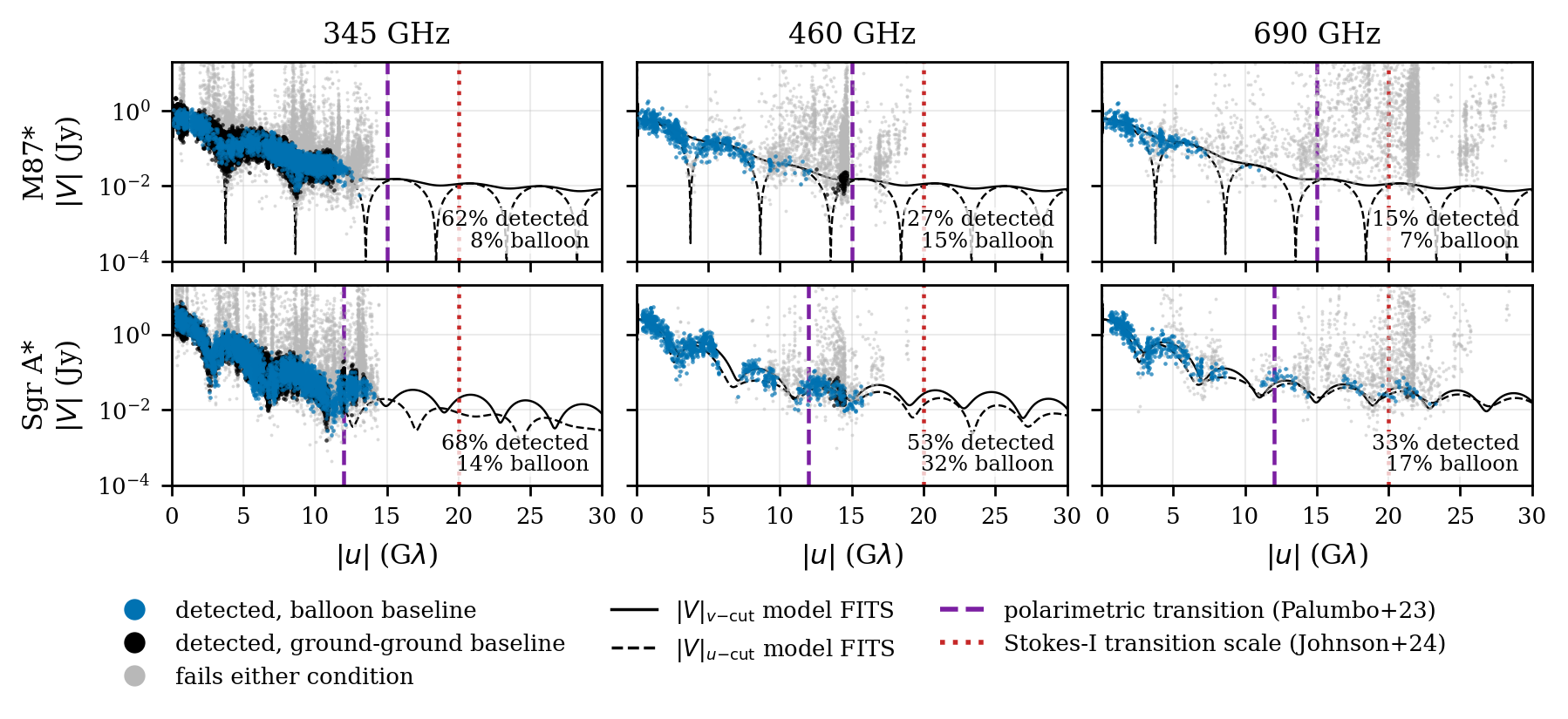}
\caption{Visibility amplitude versus baseline length for the full
arrays, assuming that frequency phase transfer can be used to obtain
300\,s scans from 5\,s (ground--balloon) or 10\,s (ground--ground)
fringe solutions on a 230\,GHz reference, over a ten-day campaign, for
M87$^*$ (top) and Sgr~A$^*$ (bottom), each with two staggered Wanaka
launches, at 345, 460, and 690\,GHz. Dark points pass the 3$\sigma$
detection rule of Section~\ref{sec:single} on balloon baselines (blue)
and ground--ground baselines (black). Gray points fail one or both of
the detection rule conditions. Solid and dashed black curves are the
$v$- and $u$-axis cuts through the model. Each panel gives the
percentage of the scheduled scans that are detected and the percentage
detected on balloon baselines. Dashed and dotted vertical lines indicate
the polarimetric transition \citep{Palumbo2023_polarimetry} and the
Stokes-I transition scale \citep{Johnson2024_BHEX}, respectively.}
\label{fig:fpt_radplot}
\end{figure*}

The thermal noise for a baseline will decrease as the dish diameter
increases, or as the integration time--frequency bandwidth product
increases, as shown in Figure~\ref{fig:sb_feasibility}.
Table~\ref{tab:treq} lists the integration time required to achieve a
$>3\sigma$ detection of Sgr~A$^*$ at 460\,GHz in median weather
conditions for different combinations of dish diameters and frequency
bandwidths. At the fiducial 32\,GHz bandwidth and 2.5\,m dish diameter
we would require an approximately 20\,s integration, which is why
Sgr~A$^*$ at 460\,GHz is at threshold at 10--20\,s. Doubling the
bandwidth halves the required integration time. A 6\,m dish would
require 3.4\,s at the fiducial bandwidth; this would be inside the
expected OCXO coherence time with no FPT needed at all.
M87$^*$ at 460\,GHz would require about 200\,s integration assuming a
2.5\,m dish, beyond the OCXO coherence but potentially achievable with
a 300\,s scan if FPT can be used.
With the band-averaged noise these requirements become 7.7\,s for the
6\,m dish, beyond the OCXO coherence, and about 470\,s for M87$^*$, longer than one scan.
At 690\,GHz the M87$^*$ requirement exceeds a single 300\,s scan at
every diameter from 2 to 6\,m. Sgr~A$^*$ would require about 440\,s
with a 2.5\,m dish, 304\,s with a 3\,m dish, just beyond a single
scan, and 76\,s with a 6\,m dish.

\subsection{Full-array coverage with frequency phase transfer} \label{sec:access}

Figure~\ref{fig:fpt_radplot} classifies every 300\,s phase-transferred
scan of the ten-day two-balloon campaigns on the full arrays
with the detection rule. Of the scheduled M87$^*$ scans, 62\%, 27\%, and 15\%
are detected at 345, 460, and 690\,GHz, and of the Sgr~A$^*$
scans 68\%, 53\%, and 33\% are detected.

The two-balloon campaigns are two launches from Wanaka staggered by
16\,d. We tested offsets from 2 to 28\,d, and 16\,d is the offset for
which the largest fraction of the scheduled 460\,GHz balloon scans lies
beyond the polarimetric transition \citep{Palumbo2023_polarimetry}.
Wanaka superpressure balloon flights, in which a
sealed balloon holds a fixed volume and floats at nearly constant
altitude for weeks, launch in March--May. SuperBIT launched there on
2023 April 16 \citep{Gill2024_SuperBIT}, inside the April window we
simulate, which is also when the EHT observes. An Antarctic long-duration flight would give the
longest daily Sgr~A$^*$ track, about 11.5 hours with ALMA, but its
December--January launch opportunities \citep{Gregory2010,Geach2021} place
Sgr~A$^*$ near the Sun. In January the Chilean weather is also at its worst,
and the median 10\,s noise on the ALMA--submm-BVEX baseline at 460\,GHz is
115\,mJy compared to 17.5\,mJy in April, so we did not simulate that
flight trajectory.

The polarimetric transition indicated with a dashed vertical line in
Figure~\ref{fig:fpt_radplot} is
the baseline beyond which the $n{=}1$ ring dominates the polarized
visibility, about 15\,\Glam{} for M87$^*$ and 12\,\Glam{} for
Sgr~A$^*$ in the 230--690\,GHz simulations of
\citet{Palumbo2023_polarimetry}. The transition moves inward with
frequency (their Figures 2 and 3). For Sgr~A$^*$ the 12\,\Glam{} value
is quoted at 345\,GHz, so we can conservatively take the transition to
be at 12\,\Glam{} at 460 and 690\,GHz. For M87$^*$ the 15\,\Glam{}
value is calculated near 690\,GHz (their Figure 4). In the Stokes-I
distribution the photon ring dominates the total intensity on baselines
longer than about 20\,\Glam{} \citep{Johnson2024_BHEX}, which we call
the Stokes-I transition scale. We use these same published values for
the 460 and 690\,GHz simulations.

For Sgr~A$^*$ the balloon stations extend the longest detected
baseline. At 460\,GHz the ground array alone reaches 14.5\,\Glam{} and
the submm-BVEX stations extend that to 16.3\,\Glam. At 690\,GHz the
ground detections are only on the 2.6\,km ALMA--APEX baseline, apart
from a few ALMA--JCMT scans (0.2\% of the scheduled scans); ALMA--APEX
is the baseline of the first 690\,GHz VLBI fringe detection
\citep{Chen2026_690GHz}. The balloon baselines provide the rest of the long-baseline
coverage out to 23.3\,\Glam. At 460\,GHz, 41\% of the ALMA--balloon
scans of the Sgr~A$^*$ campaign that pass the 230\,GHz reference
requirement (the calibratable scans) lie beyond the polarimetric
transition. Of the scheduled scans, 7.6\% at 460\,GHz and
2.3\% at 690\,GHz are detected beyond the transition, most of them on
balloon baselines. Raising the detection threshold from $3\sigma$ to $5\sigma$
keeps about three quarters of these balloon detections at 460\,GHz and
a third at 690\,GHz.

The simulations treat each phase-transferred scan as perfectly coherent,
so we also test the effect of an error in the transferred phase. The
simulation requires a reference fringe signal-to-noise ratio of 5, which
gives a thermal phase error of about 0.2\,rad per solution. We
test a larger residual reference phase error of 0.3\,rad at 230\,GHz, which
corresponds to a signal-to-noise ratio of about 3, for which the
frequency ratio scales the error to 0.6\,rad at 460\,GHz and 0.9\,rad at
690\,GHz. This error lowers the coherent visibility amplitude of a scan
by a factor of $e^{-\phi^2/2}$ for a phase error $\phi$, which is 0.84
at 460\,GHz and 0.67 at 690\,GHz, and we apply the detection rule again
to the lowered amplitudes. With this error, 93\% and 38\% of the
balloon detections beyond the transition remain. The 690\,GHz detections
beyond the transition therefore require FPT that stays
coherent at three times the reference frequency. Only 1\% of the scheduled Sgr~A$^*$ scans at
690\,GHz are detected beyond the Stokes-I transition scale, which
remains a target best suited for space VLBI \citep{Johnson2024_BHEX}.
For M87$^*$ at 460\,GHz the two Wanaka-launched balloon stations add $(u,v)$
coverage but do not extend the longest detected baseline. The ALMA--balloon baseline to M87$^*$ from
Wanaka has a median length of 7.2\,\Glam{} at 460\,GHz, so the longest
detected baseline stays the 14.7\,\Glam{} ground baseline there and
no M87$^*$ detection lies beyond the 15\,\Glam{} polarimetric
transition. At 690\,GHz the only M87$^*$ ground detections are on ALMA--APEX, so
the balloon baselines, detected out to 10.9\,\Glam, set the longest
detected baseline.

%% ============================================================================
%% SECTION 4: PHOTON RING FORECASTS
%% ============================================================================
\section{Photon Ring Forecasts} \label{sec:geometry}

\subsection{Ring model and Fisher forecast} \label{sec:model}

\begin{figure*}[!htb]
\centering
\includegraphics[width=0.90\textwidth]{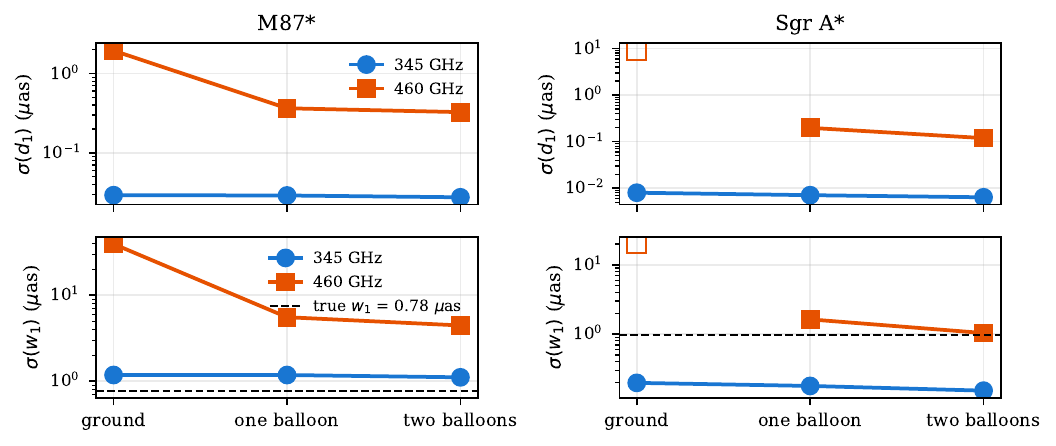}
\caption{Fisher forecast for M87$^*$ (left) and Sgr~A$^*$ (right):
$\sigma(d_1)$ (top) and $\sigma(w_1)$ (bottom, dashed line at the true
$w_1$) at 345\,GHz (blue) and 460\,GHz (orange) for the ground array
alone, with one balloon, and with two balloons, with $F_0$ fixed and
per-scan gains (10\% amplitude, flat phase) marginalized at the truth
of Table~\ref{tab:truth}. The balloons are one Wanaka flight and two
staggered Wanaka launches for both sources. Hollow markers show the
Sgr~A$^*$ 460\,GHz ground array, whose Fisher matrix is ill-conditioned
(Section~\ref{sec:results}).}
\label{fig:fisher}
\end{figure*}

We next test the ability of a ground plus balloon array to constrain
parameters in a photon ring model. For this analysis we use a simple
geometrical model of a double thick \mring, which differs from the
source images of Figure~\ref{fig:models}. The direct $n{=}0$ image,
formed by photons that complete less than half an orbit around the
black hole, is the 5-parameter asymmetric thick \mring{} of
\citet{Johnson2020}: an
infinitesimal ring of diameter $d_0$ and total flux $F_0$, modulated by
an $m{=}1$ azimuthal Fourier mode with amplitude $|\beta_1|$ and phase
$\mathrm{arg}\,\beta_1$ and convolved with a Gaussian of FWHM $w_0$.
We also include an $n{=}1$ ring as a second concentric \mring{} tied to
the first by five parameters: the flux ratio \fone, the fractional diameter offset,
the width ratio \fw, the asymmetry-amplitude ratio, and the
asymmetry-phase offset. The width uses the \citet{Tiede2022}
reparametrization $w_1 = \fw\,w_0$ with $\fw \in (0,1)$ (their
$\epsilon_1$), which enforces
a thinner $n{=}1$ ring, so the model cannot describe the same image
with the two rings exchanged.

\begin{table}[!htb]
\centering
\footnotesize
\caption{Injected double \mring{} truth for the Fisher forecasts.}
\label{tab:truth}
\begin{tabular}{lcc}
\hline\hline
Parameter & M87$^*$ & Sgr~A$^*$ \\
\hline
$F_0$ (Jy) & 0.600 & 2.500 \\
$d_0$, $w_0$ (\uas) & 45.0, 18.0 & 56.84, 22.74 \\
$|\beta_1|$, $\mathrm{arg}\,\beta_1$ & 0.4, $-\pi/2$ & 0.3, $-\pi/2$ \\
$d_1$, $w_1$ (\uas) & 38.30, 0.778 & 48.38, 0.983 \\
\fone{} $=$ \fw & $e^{-\pi} = 0.0432$ & $e^{-\pi}$ \\
\hline
\end{tabular}

\begin{minipage}{0.95\columnwidth}
\vspace{4pt}\footnotesize
The M87$^*$ truth follows \citet{Tiede2022} (their Section 2 and
Eqs. 5--7), with the $n{=}1$ ring tied to the $n{=}0$ ring by the
$e^{-\pi}$ demagnification of a $\gamma = \pi$ Lyapunov exponent
\citep{Johnson2020}; the asymmetry amplitude of 0.4 (0.15 there) and
the absence of the ellipticity they include are our choices. The
Sgr~A$^*$ truth scales the angular sizes by the ratio of gravitational
radii, 24/19. The asymmetry ratio is 1 and the asymmetry-phase offset
0 (aligned rings).
\end{minipage}
\end{table}

The forecasts use \texttt{ngEHTforecast}, evaluated at the truth of
Table~\ref{tab:truth}, on the complex visibilities of every
calibratable 300\,s scan of the ten-day campaigns, that is, every scan
whose 230\,GHz reference reaches a signal-to-noise ratio of 5 in one
integration. The Fisher matrix is built from the derivatives
of the model visibilities with respect to the parameters, weighted by
the thermal noise, and its inverse gives the parameter uncertainties.
The scans with a low signal-to-noise ratio that fail the detection
rule therefore contribute little. Note that the forecast baselines
extend beyond the longest detected ones of Section~\ref{sec:access},
because the forecasts include these scans. In addition, the Fisher
forecasts and the detection fractions of Section~\ref{sec:detect} use
different source models, so they are not directly comparable.
Per-station, per-scan complex gains are
marginalized analytically, with $F_0$ fixed (so there are nine free parameters),
10\% amplitude gains, and unconstrained phase gains, which matches the
gains injected into the synthetic data.

We report the marginalized
uncertainties on the $n{=}1$ ring diameter $d_1$ and width $w_1$ and the
signal-to-noise ratio on the flux ratio \fone, in the parametrization
of \citet{Tiede2022}. We also fitted a single and a double \mring{} to
synthetic data generated from the double \mring{} (self-fits). These
self-fits distinguish a single ring from a double ring for both
M87$^*$ and Sgr~A$^*$ at 345 and 460\,GHz.

\subsection{Forecast Results at 345 and 460\texorpdfstring{\,}{ }GHz} \label{sec:results}

The results of our Fisher forecasts are shown in Figure~\ref{fig:fisher}.
At 345\,GHz the ground array sets the constraint and the balloons
barely change it. Adding two balloons changes
$\sigma(d_1)$ from 0.029 to 0.027\,\uas{} for M87$^*$ and from
0.0080 to 0.0064\,\uas{} for Sgr~A$^*$, while $\sigma(w_1)$ changes from 1.18
to 1.11\,\uas{} for M87$^*$ and from 0.20 to 0.15\,\uas{} for Sgr~A$^*$. The 345\,GHz constraint
is set by the sensitivity of a 13-station array, so a balloon
adds long-baseline coverage there but improves the ring-geometry
precision by only 6 to 25\%.

At 460\,GHz the ground array gives no constraint. For M87$^*$ this
four-station array returns a formal $\sigma(d_0)$ of 98\,\uas, larger
than the ring, and for Sgr~A$^*$ the Fisher matrix of the three-station
array is ill-conditioned (condition number $2.4 \times 10^{14}$), so
it gives no usable uncertainties. ALMA and APEX are 2.6\,km apart on the same
plateau, so the Sgr~A$^*$ array provides one short spacing and a single
long-baseline direction toward Hawaii, from which nine free
parameters cannot be constrained. One balloon makes the constraint
possible, with $\sigma(d_1) = 0.36\,\uas$ for M87$^*$ and
0.20\,\uas{} for Sgr~A$^*$. A second, staggered launch improves it to
0.33\,\uas{} for M87$^*$ and 0.12\,\uas{} for Sgr~A$^*$, and for
Sgr~A$^*$ the signal-to-noise ratio on the flux ratio, SNR($\fone$), rises from
14 with one balloon to 24 with two (from 7 to 8 for M87$^*$). We therefore conclude that balloon
baselines are necessary for 460\,GHz constraints but would
add little additional information to the much larger 345\,GHz ground array.

For the ring geometry, the 460\,GHz forecasts do not improve on the
345\,GHz forecasts. The best
460\,GHz balloon campaign gives $\sigma(d_1)$ of 0.33\,\uas{} for
M87$^*$ and 0.12\,\uas{} for Sgr~A$^*$, against 0.029 and
0.0080\,\uas{} for the 345\,GHz ground array alone, a factor of 11 to
15 worse, and $\sigma(w_1)$ is worse by a factor of about 4 to 5. At 460\,GHz the
constraint comes from three or four ground stations plus balloons
instead of 13 ground stations. Averaging the ground-station noise over
the 460\,GHz band (Section~\ref{sec:synth}) would raise the 460\,GHz
uncertainties by a further factor of about 1.5. We do not present
ring-geometry forecast results for 690\,GHz, where the balloon detections
beyond the transition depend on FPT that stays coherent at
three times the reference frequency. Without balloons the 690\,GHz double \mring{}
Fisher matrix is singular for Sgr~A$^*$ and ill-conditioned for
M87$^*$.

%============================================================================
%% SECTION 5: DISCUSSION
%============================================================================
\section{Discussion} \label{sec:discussion}

\subsection{Access to the Polarimetric Photon Ring Regime} \label{sec:disc_pol}

Beyond the transition baseline the $n{=}1$ ring shifts the phase of the
interferometric polarization quotient of \citet{Palumbo2023_polarimetry}.
The quotient removes most station-gain terms
and cancels the diffractive scattering of Sgr~A$^*$, and it can be
averaged coherently across epochs. For Sgr~A$^*$ interstellar
scattering permits the measurement at 345\,GHz and above
\citep{Palumbo2023_polarimetry}. For M87$^*$ the transition is near 15\,\Glam,
which lies beyond the 9.8\,\Glam{} Earth diameter at 230\,GHz, so at that
frequency the measurement needs space VLBI
\citep{Palumbo2023_polarimetry}, such as BHEX \citep{Johnson2024_BHEX}. At 345\,GHz the longest M87$^*$ detection
in Figure~\ref{fig:fpt_radplot} is at 13.1\,\Glam, short of the
transition. The best target to resolve the polarimetric transition
is therefore Sgr~A$^*$, and Figure~\ref{fig:fpt_radplot} shows that both
the ground--balloon and the ground--ground baselines can exceed the
12\,\Glam{} transition at 345\,GHz, while at 460\,GHz most detections
beyond the polarimetric transition are on balloon baselines. We note
that balloon telescopes improve the long-baseline coverage by being able
to observe at higher frequencies than most ground site locations, rather
than by adding baselines longer than the Earth's diameter. For Sgr~A$^*$ in the two-balloon Wanaka
campaign, 41\% of the calibratable ALMA--balloon scans at 460\,GHz lie
beyond the 12\,\Glam{} transition (Section~\ref{sec:access}), and 61\%
at 690\,GHz.

\begin{figure}[!htb]
\centering
\includegraphics[width=\columnwidth]{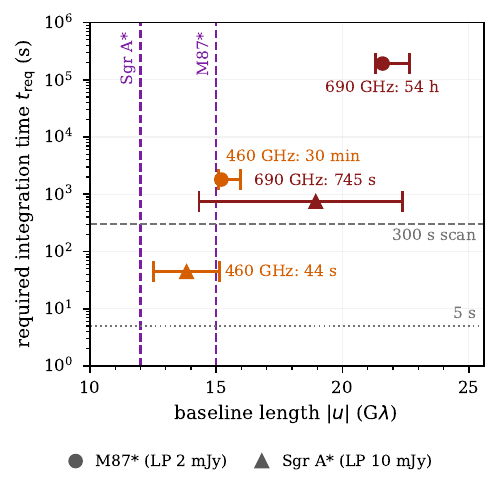}
\caption{Integration time required to reach the linear-polarization (LP)
thresholds of \citet{Palumbo2023_polarimetry}, 2\,mJy for M87$^*$
(circles) and 10\,mJy for Sgr~A$^*$ (triangles), versus baseline length
at 460\,GHz (orange) and 690\,GHz (dark red). Each point uses only the
calibratable ALMA--submm-BVEX scans beyond the polarimetric transition,
from the 3.6 days of the recorded Kiruna flight for M87$^*$ and from
both balloons of the ten-day two-balloon Wanaka campaign for Sgr~A$^*$. Markers
sit at the median baseline length of those scans, and whiskers span
their 16--84\% range. Dashed vertical lines are the polarimetric
transitions at 15 (M87$^*$) and 12\,\Glam{} (Sgr~A$^*$); the horizontal
lines are the 5\,s OCXO coherence time assumed for ground--balloon
baselines and the assumed length of a phase-transferred scan (300\,s).}
\label{fig:pol_treq_baseline}
\end{figure}

The published requirements for measuring the $n{=}1$ phase of the
polarization quotient in a single observation are a maximum
per-visibility thermal noise of about 2\,mJy for the M87$^*$
linear-polarization (LP) signature at 460--690\,GHz and about 10\,mJy
for Sgr~A$^*$ \citep{Palumbo2023_polarimetry}. The
circular-polarization (CP) requirement for M87$^*$ at 230\,GHz is more
stringent still, below 1\,mJy \citep{TamarPalumbo2024}.
Figure~\ref{fig:pol_treq_baseline} shows the integration time at which
our simulated thermal noise on the ALMA--balloon scans beyond the
transition reaches those thresholds.

For Sgr~A$^*$ the requirement is 44\,s at 460\,GHz, which is less than
our assumed length of a phase-transferred scan (300\,s), and 745\,s at
690\,GHz, about two and a half scans. ALMA--JCMT, the more sensitive of the two
ground baselines beyond the transition at 460\,GHz (the other is
APEX--JCMT, which has nearly the same length), requires about 2900\,s to
reach the 10\,mJy sensitivity needed to detect the polarimetric
transition. With the band-averaged noise at 460\,GHz, the 44\,s
requirement rises to about 100\,s, which is still less than our
assumed scan length. At 345\,GHz the ngEHT ground array reaches the
12\,\Glam{} transition by itself, and its baseline from the
Submillimeter Telescope (SMT) to the SPT reaches the LP threshold in
154\,s. However, balloon stations can make an important contribution to
this measurement at 460 and 690\,GHz, where the ground arrays reach the
transition baseline length only for the baselines from ALMA and APEX to
the JCMT.

For M87$^*$, 19\% of the calibratable ALMA--submm-BVEX scans over the
3.6 days of the recorded Kiruna flight lie beyond the transition at
460\,GHz. On the scans beyond the transition, the requirement is
30\,min at 460\,GHz and 54\,h at 690\,GHz. In the two-balloon
Wanaka campaign no ALMA--balloon scan of M87$^*$ reaches the transition
at 460\,GHz, and the most sensitive ground--balloon baseline beyond it would require
more than a hundred 300\,s scans. The CP signal, which needs a noise
below 1\,mJy, is out of reach for a 2.5\,m dish, since the median
thermal noise of two averaged 300\,s scans beyond the transition is
2.7--3.5\,mJy at 460\,GHz and 11--36\,mJy at 690\,GHz. We have not
simulated the polarized signal, so we compare the thermal noise with
published thresholds rather than make a polarimetric forecast, which
is beyond the scope of this paper.
Nevertheless, we find that balloon telescopes can provide baselines
that probe the scales beyond the polarimetric transition, where the
$n{=}1$ ring of Sgr~A$^*$ dominates the polarized visibility,
especially the ALMA--submm-BVEX baselines at 460\,GHz, and possibly at
690\,GHz if the scans can be coherently averaged.

\subsection{Limitations} \label{sec:disc_limits}

In our simulations we assume a roughly 5\,s balloon coherence time,
limited by the OCXO timing stability and position tracking system
precision, and phase-connected 230\,GHz reference solutions. The
simulation treats every 300\,s phase-transferred average as coherent.
The reference 230\,GHz frequency band is recorded at the science
bandwidth. We also find that detections beyond the polarimetric
transition baseline at 690\,GHz will depend on whether we can
coherently average the 5\,s integrations over a 300\,s scan
(Section~\ref{sec:access}).

Our atmospheric opacity calculations for the ground stations assume
that the opacity at the center of the band applies across the entire
band. This assumption underestimates the opacity in the 460\,GHz band,
and consequently underestimates the thermal noise by a factor of about
1.5, because the band contains the 448\,GHz water line
(Section~\ref{sec:synth}). In future forecasting work we could mitigate
this higher ground-station noise by either increasing the central
frequency of the band or tuning the two sidebands of the 2SB receivers
to two 16\,GHz windows with better atmospheric transmission properties.

The Fisher forecasts presented in this paper marginalize over a 10\%
amplitude and a free phase gain per station and scan, and non-closing
errors such as bandpass mismatch and polarization leakage are not
included in our analysis. The Fisher truth includes no interstellar
scattering, whereas the Sgr~A$^*$ model images used for the simulations
of Section~\ref{sec:detect} do include scattering. Finally, the
analysis assumes that the sources do not change over the ten-day
campaign. This is not true for Sgr~A$^*$, which varies on timescales
of minutes \citep{EHTC2022_IV}, so the forecasts overstate what a
single Sgr~A$^*$ epoch gives.

In contrast, the analysis of Section~\ref{sec:single} treats one
baseline with no FPT, fringe grouping, or averaging across
integrations, which gives a conservative lower bound on whether fringes
can be detected on the ALMA--submm-BVEX baseline, the balloon baseline
with the lowest thermal noise. With a 2.5\,m balloon telescope paired with ALMA, both sources are
detected on more than half of their 10\,s integrations at 345\,GHz, and
Sgr~A$^*$ at 460\,GHz is at the detection threshold.

We note that the 460\,GHz receivers at the JCMT and the GLTS, the
690\,GHz receiver at the GLTS, and every submm-BVEX parameter are
targets rather than fielded hardware.

\subsection{Outlook} \label{sec:disc_outlook}

Here we discuss what would need to be done for submillimeter balloon
VLBI to be feasible. Constraining the photon ring of Sgr~A$^*$ at
460\,GHz would require a 2.5\,m balloon telescope with
dual-polarization receivers, 32\,GHz of recorded bandwidth, and FPT
from a simultaneously observed 230\,GHz band. ALMA would also need to
observe the 230 and 460\,GHz bands at the same time. This capability is
not part of the WSU \citep{Carpenter2023_WSU}.

ALMA is currently increasing its recorded bandwidth through the WSU. A
further doubling to 64\,GHz, beyond the WSU goal of 32\,GHz, would
halve the 32\,GHz integration times in Table~\ref{tab:treq} and bring the
Sgr~A$^*$ 460\,GHz requirement for the 2.5\,m dish to about 10\,s. The other upgrade is dish diameter. The 2.5\,m BLAST-TNG
primary is among the largest mirrors flown on a balloon
\citep{Lourie2018}, and 2.5--3\,m is the largest telescope size that
can be launched in the near term. A 6\,m dish (the last column of
Table~\ref{tab:treq}) brings the Sgr~A$^*$ 460\,GHz requirement to
3.4\,s, inside the expected balloon-station coherence time. With the band-averaged noise the requirement becomes
7.7\,s, longer than that coherence, and a dish that size is beyond
what a balloon can lift today.

BHEX targets orbital baselines beyond
the Stokes-I transition scale \citep{Johnson2024_BHEX,Lupsasca2024},
whereas a balloon reaches the polarimetric regime of Sgr~A$^*$ at 460
and 690\,GHz, at shorter baselines. Balloon stations could also add
baselines to BHEX in its own observing bands, and we leave this to a
future paper.

%% ============================================================================
%% SECTION 6: CONCLUSIONS
%% ============================================================================
\section{Conclusions} \label{sec:conclusions}

In this paper we present forecasts for a submillimeter successor
experiment to the Balloon-borne VLBI Experiment (BVEX). We assume a
2.5\,m telescope at 35--40\,km that has effectively no atmospheric
contribution to its noise, so it can observe at 460 and 690\,GHz from
any latitude or longitude. Ground VLBI at those frequencies is confined
to a few dry sites, and balloon-borne VLBI stations can improve the
$(u,v)$ coverage at these frequencies, where a fixed baseline is longer
in wavelengths.

The pathways we identify for balloon telescopes to help ground-based
arrays constrain photon ring models differ between 345, 460, and
690\,GHz. For all submillimeter bands we assume simultaneous
observations at 230\,GHz as an FPT reference band, to obtain longer
coherence times in the higher-frequency science bands. On the single
ALMA--balloon baseline the 230\,GHz reference fringe is detected on
every 10\,s integration for M87$^*$ and on about half for Sgr~A$^*$,
whose scattered image has a visibility amplitude only a few times the
noise on this baseline.
At 345\,GHz the ngEHT ground array primarily sets the ring-geometry
precision, though two balloons add baselines at and beyond the longest
ngEHT spacings and improve that precision by 6 to 25\%. At 460\,GHz,
Sgr~A$^*$ is near the detection threshold on the ALMA--submm-BVEX
baseline, with ten or more 20\,s detections on about half of
April days for a 2.5\,m dish at the WSU goal bandwidth, while M87$^*$
would require FPT to make significant detections. A 460\,GHz
ring-geometry constraint exists only with balloons, though it would be
less precise than the constraint from the 345\,GHz ground array alone
because of the smaller number of VLBI stations. At 690\,GHz neither
source is detected on the single baseline at any balloon telescope
diameter up to 6\,m or bandwidth up to 64\,GHz without FPT.

The most interesting potential science case for submillimeter balloon
VLBI is the polarimetric photon ring regime.
The single-baseline Sgr~A$^*$ detections at 460\,GHz lie beyond the
polarimetric transition of \citet{Palumbo2023_polarimetry}. With FPT, most
of the Sgr~A$^*$ detections beyond the transition at
460\,GHz are on balloon baselines, and the ground array reaches the
transition there only on the baselines from ALMA and APEX to the JCMT.
At 690\,GHz such detections would require
FPT that stays coherent at three times the reference
frequency. No M87$^*$ detection on the balloon baselines from Wanaka
reaches the transition baseline length in our forecasts. For Sgr~A$^*$ the thermal noise on the ALMA--balloon scans
beyond the transition meets the published linear-polarization threshold
within one 300\,s scan at 460\,GHz.

The steps toward a submillimeter balloon VLBI experiment are a 32\,GHz bandwidth backend, the
submillimeter receivers of Table~\ref{tab:stations}, simultaneous
recording of the 230\,GHz reference at every station, and multi-launch
campaigns from one site. After balloon-borne VLBI is demonstrated, a
next step toward such an experiment could be a demonstration of FPT applied to
balloon VLBI observations at lower frequencies (e.g., 22 and 88\,GHz).

\FloatBarrier
\begin{acknowledgments}
The authors would like to thank the Centre National d'\'Etudes Spatiales
(CNES) for providing GPS data of the 2024 TRANSAT flight from Kiruna used
for the analysis in this paper through collaboration with the Canadian
Space Agency (CSA).

MB, FMT and LMF would like to acknowledge the support from the Canadian Space
Agency (CSA) Flights and Fieldwork for the Advancement of Science and
Technology (FAST) grants (21FAQUEA18 and 25FAQUB72) as well as the support from the NSERC
Collaborative Research and Training Experience (CREATE) Radio
Astronomy-Driven Education and Training Excellence (RADEATE) program
(CREATE 596358-2025). This research was also supported by the Canada Foundation for Innovation (CFI) John R. Evans
Leaders Fund (JELF Project Number 43766), the Natural Sciences and Engineering Research Council
(NSERC)  Discovery Grant RGPIN/06266-2020 (LMF), and
a Queen's University Research Initiation
Grant.
\end{acknowledgments}

\software{\texttt{ngehtsim} \citep{Pesce2024},
\texttt{eht-imaging} \citep{Chael2016,Chael2018},
\texttt{am} \citep{Paine2023_am},
\texttt{ngEHTforecast} (\url{https://github.com/aeb/ngEHTforecast}).
This work uses versions of \texttt{ngehtsim} and \texttt{eht-imaging}
modified to support balloon stations (Section~\ref{sec:synth}).}

%% ============================================================================
%% REFERENCES
%% ============================================================================
\bibliography{references}{}
\bibliographystyle{aasjournalv7}

\end{document}